\documentclass[12pt]{article}
\let\section=\subsection     \let\subsection=\subsubsection                

\usepackage{graphicx}
\begin{document}
%
%
\begin{center}
   {\large \bf Spectroscopy of heavy nuclei by configuration mixing}\\[2mm]
   {\large \bf of symmetry restored mean-field states:}\\[2mm]
   {\large \bf Shape coexistence in neutron-deficient Pb isotopes}\\[3mm]
    M.~Bender,$^{1}$ P.~Bonche,$^{2}$ T.~Duguet,$^{3}$ and 
    P.-H.~Heenen$^{1}$\\[3mm]
   {\small \it $({1})$ Service de Physique Nucl\'eaire Th\'eorique
    et de Physique Math{\'e}matique,\\
    Universit\'e Libre de Bruxelles, Belgium\\
    $({2})$ Service de Physique Th\'eorique,
    CEA Saclay, Gif sur Yvette Cedex, France\\
    $({3})$ Physics Division, Argonne National Laboratory,
    Argonne, IL 60439, U.S.A.\\[4mm]}
\end{center}
%
%
\begin{abstract}\noindent
We study shape coexistence and low-energy excitation spectra
in neutron-deficient Pb isotopes using configuration mixing of
angular-momentum and particle-number projected self-consistent
mean-field states. The same Skyrme interaction SLy6 is used everywhere
in connection with a density-dependent zero-range pairing force.
\end{abstract}
%
%
Methods well suited to describe microscopically medium
and heavy nuclei are based on effective energy functionals
and self-consistent mean-field models \cite{RMP}. They 
are successful for the description of many nuclear properties, but do 
not give direct access to low-energy spectroscopy, which in many cases
requires to include correlations beyond the mean field approach. 
Broken symmetries can be restored by projection. Particle-number projection 
removes spurious contributions coming from states with different 
particle numbers, which are an artifact of the usual BCS approach. 
Angular momentum projection restores rotational symmetry and generates 
wave functions in the laboratory frame which provide transition 
probabilities and spectroscopic moments without further approximations.
Finally, the variational configuration mixing with respect to a
collective coordinate by means of the Generator-Coordinate Method
(GCM) allows for an improved calculation of nuclear ground-state 
properties, as it removes the contributions from spurious collective 
vibrations that are inevitably mixed to the mean-field states.  
Besides that, a GCM calculation provides the excitation spectrum 
corresponding to given collective coordinates. 

A model that combines these extensions of the mean-field approach 
offers a powerful tool: from a numerical point-of-view it is still 
simple enough to be applied to all up to superheavy 
nuclei, using the full model space of single-particle states with 
realistic asymptotics and the proper coupling to the continuum. 
Correlations corresponding to collective modes can be incorporated
step by step into the modeling which helps to identify the relevant 
degrees of freedom. The method has the advantage that its results can 
be interpreted within the intuitive picture of intrinsic shapes 
and shells of single-particle states that is offered by the framework 
of mean-field models.

We present here an application of a method based on these principles
to neutron-deficient Pb isotopes around $^{186}$Pb. The ground 
state of Pb isotopes 
is known to be spherical down to $^{182}$Pb. However, weakening 
of the \mbox{$Z=82$} shell manifests itself by the appearance 
of low-lying $0^+$ states \cite{Jul01}. At least one low-lying 
excited $0^+$ state has been observed in all even-even Pb isotopes 
between $^{182}$Pb and $^{194}$Pb at excitation energies 
below 1 MeV, the most extreme cases being $^{186}$Pb \cite{And00} 
with two excited $0^+$ states below 700 keV. 
%
%
\subsubsection*{The Method}
The starting point of our method is a set of HF+BCS wave functions 
$| q \rangle$ generated by self-consistent mean-field calculations 
with a constraint on a collective coordinate $q$. Such mean-field 
states break several symmetries of the exact many-body states.
Wave functions with good angular momentum and particle numbers 
are obtained by restoration of rotational and particle-number 
symmetry on $ | q \rangle$:
\begin{equation}
\label{eq:proj}
|J M q \rangle 
= \frac{1}{{\mathcal N}}
  \sum_{K} g^{J}_{K}
  \hat{P}^J_{MK} \hat{P}_Z \hat{P}_N | q \rangle
,
\end{equation}
where ${\mathcal N}$ is a normalization factor. 
$\hat{P}^{J}_{MK}$, $\hat{P}_N$, $\hat{P}_Z$ are projectors 
onto angular momentum $J$ with projection $M$ along the
laboratory z-axis, neutron number $N$ and proton number $Z$ respectively. 
We impose here axial symmetry and time reversal invariance
on the intrinsic states $| q \rangle$. Therefore, $K$ can only be 
0 and we shall omit the coefficient \mbox{$g^{J}_{K} = \delta_{K0}$}.
A variational configuration mixing on the collective variable 
$q$ is then performed for each $J$
\begin{equation}
\label{eq:discsum}
| J M k \rangle 
= \sum_{q} f_{k}^{JM} (q) | JM q \rangle 
.
\end{equation}
The weight functions $f_k^{JM}(q)$ are determined by requiring
that the expectation value of the energy
\begin{equation}
\label{eq:egcm}
E^{JM}_k
= \frac{\langle JM k | \hat H | JM k \rangle}
       {\langle JM k | JM k\rangle} ,
\end{equation}
is stationary with respect to an arbitrary variation $\delta f_k^{JM}(q)$.
This prescription leads to the discretized Hill-Wheeler equation. 
Collective wave functions in the basis of 
the intrinsic states are then obtained from the set of weight 
functions $f_k^{JM}(q)$ by a basis transformation \cite{Taj93}.
In $| J M k \rangle$, the weight of each mean-field state $| q \rangle$
is given by:
\begin{equation}
\label{eq:weight}
g^{JM}_k(q)
= {\langle JM k | q \rangle}\quad.
\end{equation}
Since the collective states $| JM k \rangle$ have good angular 
momentum, quadrupole moments and transition probabilities can 
be directly determined in the laboratory frame of reference without
further approximations. For further details on the evaluation of
overlaps and matrix elements see \cite{Val01}.

The same effective interaction is used to generate the mean-field wave 
functions and for the configuration mixing calculation. We have chosen 
the Skyrme interaction SLy6 in the mean-field channel \cite{Cha98} 
and a density-dependent zero-range force as defined in \cite{Rig99}
in the pairing channel. 
%
%
\subsubsection*{An illustrative example: $^{186}$Pb}
%
%
\begin{figure}[t!]
\centerline{\includegraphics{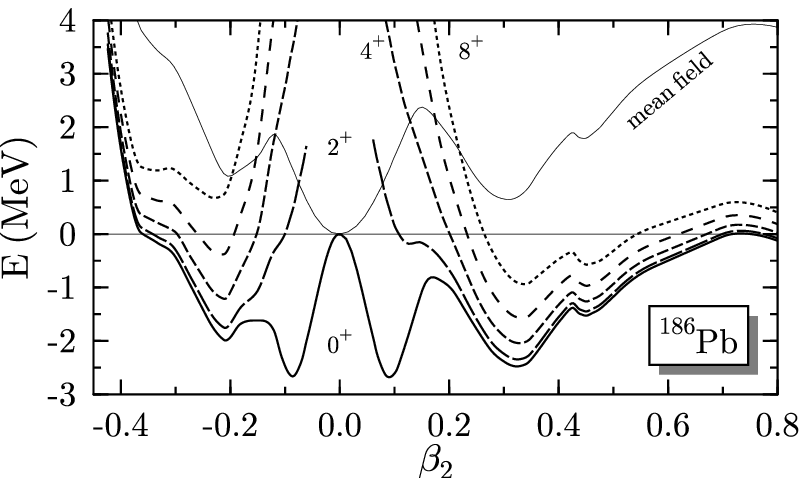}}
{\centerline{\footnotesize\parbox{14cm}{
Figure 1: Particle-number projected (``mean field'') and particle-number
and angular-momentum projected potential energy curves up to
\mbox{$J=10$} for $^{186}$Pb as a function of the mass
quadrupole deformation $\beta_2$. 
}}}
\end{figure}
%
%
We discuss $^{186}$Pb as an example for the results that can be 
obtained with the method; see ref.\ \cite{Dug03} for more details. 
On figure~1 is plotted the deformation energy of 
$^{186}$Pb before and after projection on angular momentum.
All curves are drawn versus the intrinsic axial quadrupole moment 
of the unprojected mean-field states. As projected \mbox{$J=0$} 
states are spherical, this ``quadrupole moment'' is only a 
convenient way to label the projected states. The curve labeled 
``mean-field'' plots the deformation energy after particle-number 
projection only. It exhibits a spherical global minimum as well 
as local minima at prolate and oblate deformations.
While the deformation energy of the prolate minimum fortuitously
reproduces the experimental value of 0.650 MeV for the prolate 
$0^+$ state, the 1.1 MeV deformation energy of the oblate minimum 
overestimates the experimental value of 0.532 MeV for 
the oblate $0^+$ state. 
%
%
\begin{figure}[t!]
\centerline{\includegraphics{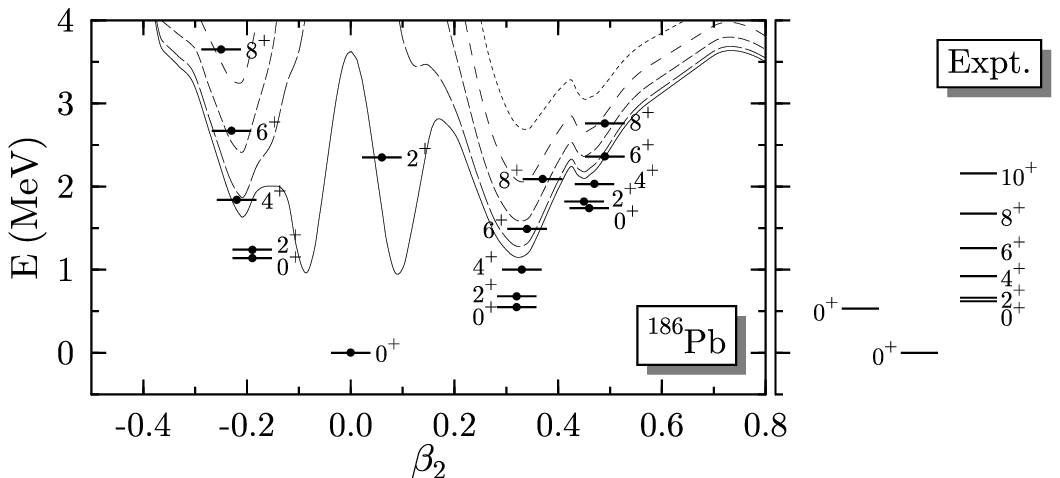}}
{\centerline{\footnotesize\parbox{14cm}{
Figure 2:
Left panel: Spectrum of the lowest positive parity bands with even angular 
momentum and \mbox{$K=0$}, as a function of the deformation (see text).
The angular momentum projected energy curves are shown for comparison.
The energy reference is that of the calculated $0^+_1$ ground state. 
Right panel: Available experimental excitation energies 
\cite{And00,Hee93}. From the left to the right the spectrum shows 
oblate, spherical and prolate states.
}}}
\end{figure}
%
%
%
\begin{figure}[b!]
\centerline{\includegraphics{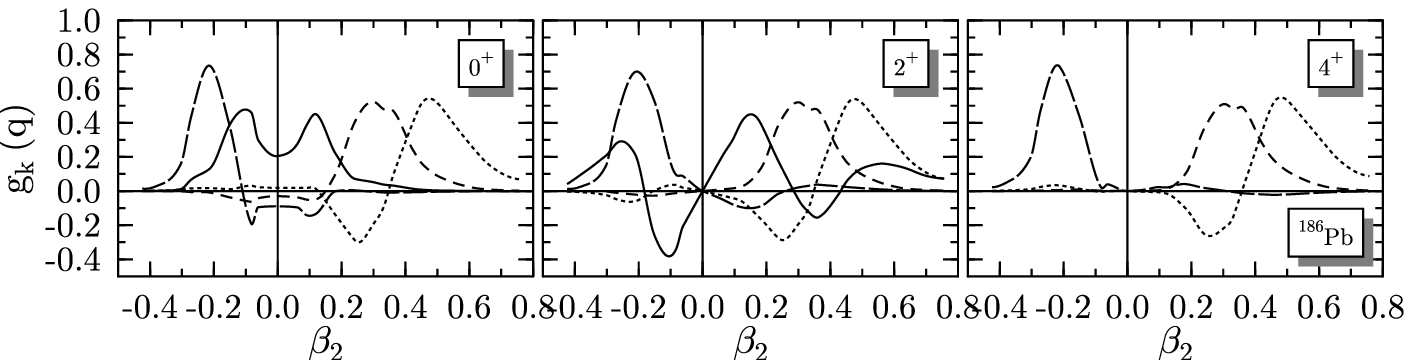}}
{\centerline{\footnotesize\parbox{14cm}{
Figure 3: GCM wave functions of the lowest $| J k \rangle$ states. 
Solid lines denote spherical, long-dashed lines oblate, dashed lines 
prolate, dotted lines the $\beta$ band in the prolate well.
}}}
\end{figure}
%
%

The energy curves obtained after angular momentum projection are also 
shown on figure~1. The spherical mean-field state
is rotationally invariant and therefore contributes to \mbox{$J=0$} 
only. Two minima appear at small deformations around 
\mbox{$\beta_2 = \pm 0.1$}. They do not correspond to two different 
states, but to the correlated spherical state (see below). 

The excitation energies $E^{JM}_k$ of the collective states 
$| JM k \rangle$ obtained from the configuration mixing calculation
are shown in figure~2. Each of these states is 
represented by a bar drawn at the average intrinsic 
deformation of the mean-field states it is built from.
The excitation spectrum is divided into bands corresponding to 
different deformations. Configuration mixing lowers the energy of 
the lowest collective states with respect to the projected energy 
curves; the energy gain is the largest for the ground state. 
The available data are plotted on the right panel of figure~2. 
The calculated energy of the prolate $0^+$ state 
is very close to the experimental energy contrary to the excitation 
energy of the oblate $0^+$ state which is largely overestimated. 

The corresponding collective wave functions $g^{JM}_k(q)$ are 
plotted in figure~3. 
The ground state wave-function is spread in a similar way on both 
oblate and prolate sides with a zero average quadrupole deformation.
The wave functions of the first two excited $0^+$ states are 
strongly peaked at either prolate or oblate deformations, with their 
tails extending into the spherical well. Starting with 
\mbox{$J=4$}, states are localized and are predominantly 
either prolate or oblate.
The shapes of the $0^+_4$, $2^+_3$, and $4^+_3$ wave functions suggests 
their interpretation as a rotational band built onto a $\beta$ vibration 
within the prolate well.
%
%
\begin{figure}[t!]
\centerline{\includegraphics{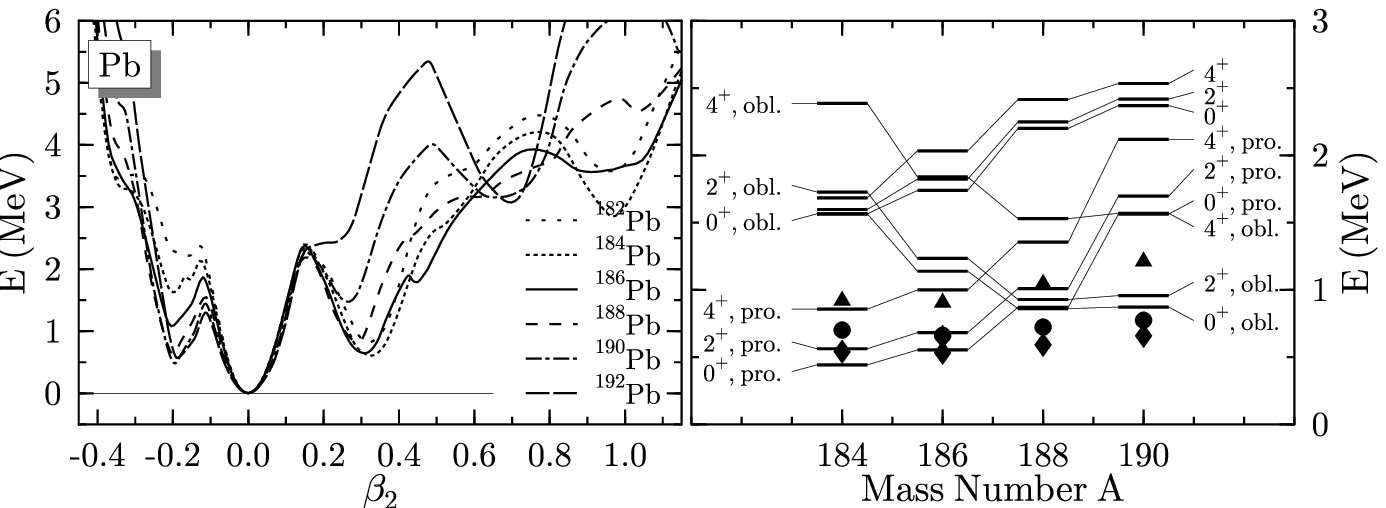}}
{\centerline{\footnotesize\parbox{14cm}{
Figure 4: Left panel: mean-field potential energy curves for 
{$^{182-192}$Pb}.
Right panel: Systematics of low-lying states. The two lowest 
calculated $0^+$ and $2^+$ states in $^{188}$Pb are nearly degenerate.
Filled diamonds, circles, and triangles denote the lowest known excited
$0^+$, $2^+$, and $4^+$ states respectively. Data taken from 
\protect\cite{Jul01}.
}}}
\end{figure}
%
%

Calculated transition probabilities for \mbox{$J>2$} states confirm 
the separation of the excited states into rotational bands with very small 
$B(E2)$ transitions between bands, while the large branching of $E2$ 
strength for the in and out of bands $2^+ \to 0^+$ transitions reflects 
that the low-lying $0^+$ states are mixed. 
%
%
\subsubsection*{Systematics}
The systematics of the potential energy surfaces for the Pb isotopes
with \mbox{$A=182-192$} is shown in the left panel of figure~4. 
It can be clearly seen that the oblate minimum is pushed up in energy and 
disappears below \mbox{$A=184$}, while a prolate minimum develops
below \mbox{$A=192$}. Various superdeformed structures can be seen
at large deformation. The pattern is reflected in the systematics of
the excitation energies of the $0^+$, $2^+$ and $4^+$ states in 
the three lowest rotational bands obtained from the configuration 
mixing for \mbox{$A=184-190$}, see the right panel of figure~4.
Calculated oblate and prolate bands cross for $^{188}$Pb, at slightly
larger neutron number than suggested by experiment \cite{Jul01}.
For the three heavier isotopes, the third band has a similar 
structure as the one obtained for $^{186}$Pb, while for $^{184}$Pb 
it corresponds to the superdeformed minimum around 
\mbox{$\beta_2 = 1.0$}.
%
%
\subsubsection*{Conclusions}
Systematic calculations of spectroscopic properties of very heavy nuclei 
based on self-consistent mean-field methods have become feasible,
as has been shown on the example of low-lying excited states in
neutron-deficient Pb isotopes.
Our results strongly support the interpretation of the Pb spectra 
as a manifestation of shape coexistence. There remains, however,
a significant over-estimation of the oblate band's excitation energy
in some of the nuclei, which can be due to the one or the other ingredient 
of the model: (1) the effective mean-field interaction (as surface 
tension, spin-orbit splitting, etc), (2) the strength and the form 
factor of the pairing interaction, (3) missing collective modes in
the configuration mixing, or (4) a necessary generalization of the
density dependence of the effective interaction for calculations 
beyond mean field as suggested in ref.\ \cite{Dug02}. The effective
interactions clearly deserve future reexamination. Various 
generalizations of the model are underway.\\[-2mm]

%
%
\noindent
\textbf{Acknowledgments.}
This research was supported in part by the \mbox{PAI-P5-07} of the 
Belgian Office for Scientific Policy, and by the US Department of Energy 
(DOE), Nuclear Physics Division, under contract no.\ W-31-109-ENG-38.
We thank M.~Huyse, R.~V.~F.\ Janssens, R.\ Julin, G.\ Neyens, and 
P.\ Van Duppen for inspiring discussions. M.~B.\ acknowledges support 
through a European Community Marie Curie \nolinebreak  Fellowship.\\[-6mm]
%
%


\begin{thebibliography}{99}

\itemsep=-0.1cm
 
\bibitem{RMP}
M.~Bender, P.-H.~Heenen, P.-G.~Reinhard,
Rev. Mod. Phys. \textbf{75} (2003) \nolinebreak 121.

\bibitem{Jul01}
R. Julin \emph{et. al.}, 
J. Phys. G \textbf{27} (2001) R109.
 
\bibitem{And00}
A. N. Andreyev \emph{et. al.},
Nature \textbf{405} (2000) 430.
 
\bibitem{Taj93}
N. Tajima \emph{et. al.}, 
Nucl. Phys. \textbf{A551} (1993) 409.
 
\bibitem{Val01}
A. Valor \emph{et. al.}, 
Nucl. Phys. \textbf{A671} (2001) 145.

\bibitem{Cha98}
E. Chabanat \emph{et. al.}, 
Nucl. Phys. \textbf{A635} (1998) 231.

\bibitem{Rig99}
C. Rigollet \emph{et. al.}, 
Phys. Rev. C \textbf{59} (1999) 3120.
 
\bibitem{Dug03}
T. Duguet \emph{et. al.}, preprint nucl-th/0212016. 

\bibitem{Hee93}
J. Heese \emph{et. al.},
Phys. Lett. \textbf{B302} (1993) 390.

\bibitem{Dug02}
T. Duguet and P. Bonche, preprint nucl-th/0210057.

\end{thebibliography}
\end{document}